\documentclass[aip,pop,reprint,a4paper]{revtex4-1}
\usepackage{cancel}
\usepackage{amsmath}
\usepackage{amssymb}
\usepackage{graphicx}
\usepackage{dcolumn}
\usepackage{bm}
\usepackage{upgreek}
\usepackage{caption}
\usepackage{subcaption}
\usepackage{comment}
\usepackage{color}

\newcommand{\meff}{\widetilde{m}}
\newcommand{\rhoi}{\overline{\varrho}_i}
\newcommand{\rhoe}{\overline{\varrho}_e}

\newcommand{\Fn}{\frac{\delta\mathcal{F}}{\delta n}}
\newcommand{\FLL}{\frac{\delta\mathcal{F}}{\delta\bm{L}}}
\newcommand{\FBB}{\frac{\delta\mathcal{F}}{\delta\bm{B}}}

\newcommand{\Frhos}{\frac{\delta\mathcal{F}}{\delta\overline{\varrho}_\sigma}}

\newcommand{\Gn}{\frac{\delta\mathcal{G}}{\delta n}}
\newcommand{\GLL}{\frac{\delta\mathcal{G}}{\delta\bm{L}}}
\newcommand{\GBB}{\frac{\delta\mathcal{G}}{\delta\bm{B}}}

\newcommand{\Grhos}{\frac{\delta\mathcal{G}}{\delta\overline{\varrho}_\sigma}}

\includecomment{comment}     

\begin{document}


\title{Hamiltonian structure of the guiding center plasma model} 



\author{J. W. Burby}
 \affiliation{Courant Institute of Mathematical Sciences, New York, New York 10012, USA}
\author{W. Sengupta}
 \affiliation{Courant Institute of Mathematical Sciences, New York, New York 10012, USA}


\date{\today}

\begin{abstract}
The guiding center plasma model (also known as kinetic MHD) is a rigorous sub-cyclotron-frequency closure of the Vlasov-Maxwell system. While the model has been known for decades, and it plays a fundamental role in describing the physics of strongly-magnetized collisionless plasmas, its Hamiltonian structure has never been found. We provide explicit expressions for the model's Poisson bracket and Hamiltonian, and thereby prove that the model is an infinite-dimensional Hamiltonian system. The bracket is derived in a manner that ensures it satisfies the Jacobi identity. We also report on several previously-unknown circulation theorems satisfied by the guiding center plasma model. Without knowledge of the Hamiltonian structure, these circulation theorems would be difficult to guess.
\end{abstract}


\maketitle 


\section{Introduction}
A problem that captivated a number of the pioneering researchers\cite{grad_review_1966} in the field of magnetized plasma physics was that of providing a rational explanation for the success of magnetohydrodynamics (MHD), especially in collisionless plasmas. Fluid models, including MHD, are most readily justified by a preponderance of collisions, which formally justifies the use of Chapman-Enskog theory to derive rigorous asymptotic fluid closures. However, when collisions are very weak, such fluid closures break down. Why then is a single-fluid model like MHD capable of describing many of the observed features of magnetized collisionless plasmas?

A great deal of insight into this problem was generated in the late 1950's and early 1960's by the arrival of what is called either the guiding center plasma (GCP) model, or kinetic MHD. (``Kinetic MHD" is the term used most frequently by astrophysicists, while ``guiding center plasma" may resonate more strongly in the magnetic fusion community.) The guiding center plasma model is a rigorous low-frequency closure of the (collisionless) Vlasov-Maxwell system that describes magnetized plasma dynamics with temporal variations much slower than the cyclotron period, spatial variations much longer than the gyroradius, and macroscopic flows that are comparable to the thermal speed. The model is \emph{not} hybrid species-wise as in the model by Cheng in Ref.\,\onlinecite{Cheng_1991}, nor is it the same as the so-called guiding center Vlasov-Maxwell model studied in Ref.\,\onlinecite{Brizard_Tronci_2016} and discretized in Ref.\,\onlinecite{Evstatiev_2014}. Perhaps most critically, the model differs substantially from gyrokinetics,\cite{Brizard_2007,Abel_2013} which allows for gyroradius-scale perpendicular wavelengths, and often assumes a low-flow ordering. The formulation of the model is due to the roughly-contemporaneous work of Chew, Goldberger, Low,\cite{CGL_1956} and Grad,\cite{Grad_1956} while further crucial insights into the model's properties were found, for example, by Kruskal and Oberman,\cite{Kruskal_Oberman_1958} Rosenbluth and Rostoker,\cite{Rosenbluth_Rostoker_1958} Kulsrud,\cite{Kulsrud_1962,Kulsrud_book_1983} Ramos,\cite{Ramos_2008,Ramos_KMHD_2015,Ramos_stab_2016} and Cerfon and Freidberg.\cite{Cerfon_2011} Fluid closures of the model were developed originally in Ref.\,\onlinecite{CGL_1956} and refined to account for parallel kinetic effects in Ref.\,\onlinecite{Snyder_1997}.
Guiding center plasma theory shows that, in the asymptotic limit on which the theory is based, the plasma density, momentum, and magnetic field are governed by MHD with a pressure tensor that is self-consistently determined by the ion and electron distribution functions. Moreover, the distribution functions are governed by so-called drift kinetic equations, which account for $E\times B$-drift, parallel streaming, magnetic mirroring, and (weak) parallel electric forcing. Thus, the guiding center plasma model shows that strongly-magnetized plasmas are \emph{nearly} governed by MHD, with the only significant departure from MHD behavior being due to kinetic effects parallel to the magnetic field. 

As one might expect of such a fundamental description of nature, the guiding center plasma model enjoys a number of exact conservation laws. (See Ref.\,\onlinecite{Ramos_stab_2016} for a particularly clean discussion.) Total energy and total momentum are neither created nor destroyed. Moreover, due to the spatially- and temporally-local nature of the asymptotic assumptions that underpin the theory, local energy and momentum conservation laws have been found. In addition, magnetic flux through any loop that is entrained in the mean fluid flow is also conserved, which is a direct consequence of the MHD Faraday equation. 

The presence of these conservation laws, together with the absence of collisions, and the rigorous connection with Vlasov-Maxwell kinetic theory, strongly suggests that the guiding center plasma model is a Hamiltonian system. However, in spite of the model's age and the attention it has garnered from various plasma physics luminaries,\cite{grad_review_1966} the guiding center plasma's Hamiltonian structure, or lack thereof, has gone undiscovered to this day. This is unfortunate due to the wide variety of theoretical tools that are applicable only to Hamiltonian systems. Examples include KAM-theoretic perturbation methods,\cite{Wayne_1990} statistical-mechanical tools such as the partition function,\cite{Bourgain_1996} structure-preserving simulation techniques,\cite{He_2016,Zhou_2014,Burby_fdk_2017,Kraus_GEMPIC_pub_2017} equilibrium variational principles and energy principles\cite{Holm_stability_1985,Morrison_deltaF_1990} (interestingly, some of these are actually known for kinetic MHD\cite{grad_review_1966,Ramos_stab_2016}), and structure-preserving model reduction techniques.\cite{Burby_two_fluid_2017}

In this Letter we will prove that, on infinite or periodic spatial domains, the guiding center plasma model is indeed an infinite-dimensional Hamiltonian system. We will do this by producing explicit expressions for the theory's Hamiltonian and Poisson bracket. Our results may apply as they stand in domains with a fixed conducting boundary, but we cannot presently confirm this. Our derivation of the bracket will prove that it satisfies the defining properties of a Poisson bracket, especially the Jacobi identity. At the end of the Letter, we will also report on several new circulation theorems satisfied by the GCP model. These circulation theorems, which are phase-space analogues of Kelvin's circulation theorem, are deeply connected to the GCP Hamiltonian structure.

\section{The guiding center plasma model and its Hamiltonian structure}
In a form close to the one given by Ramos in Ref.\,\onlinecite{Ramos_stab_2016}, the kinetic MHD (i.e. GCP) system of equations is given by
\begin{gather}
\partial_t n+\nabla\cdot \left(\frac{\bm{L}}{\meff }\right)=0\label{eq:kmhd_continuity}\\
\partial_t\bm{L}+\nabla\cdot\left(\frac{\bm{L}\bm{L}}{\meff n}+\mathbb{P}_i+\mathbb{P}_e\right)=\nabla\cdot\mathbb{S}_M\label{eq:kmhd_momentum}\\
\partial_t\bm{B}=\nabla\times\left(\frac{\bm{L}\times\bm{B}}{\meff n}\right)\label{eq:kmhd_faraday}\\
\partial_t(n\rhoi)+\nabla\cdot\left(\left[w_\parallel \bm{b}+\frac{\bm{L}}{\meff n}\right]n\rhoi\right)\nonumber\\
+\partial_{w_\parallel}\left(\left[\frac{\bm{b}\cdot(\nabla\cdot\mathbb{P}_i)}{m_i n}-\frac{\mu}{m_i}\bm{b}\cdot\nabla B \right]n\rhoi\right)\nonumber\\
-\partial_{w_\parallel}\left(\left[w_\parallel\bm{b}\bm{b}:\nabla\frac{\bm{L}}{\meff n}\right]n\rhoi\right)=0\label{eq:kmhd_ion_dke}\\
\partial_t(Z_i n \rhoe)+\nabla\cdot\left(\left[w_\parallel\bm{b}+\frac{\bm{L}}{\meff n}\right]Z_i n\rhoe\right)\nonumber\\
+\partial_{w_\parallel}\left(\left[\frac{\bm{b}\cdot(\nabla\cdot\mathbb{P}_e)}{m_e Z_i n}-\frac{\mu}{m_e}\bm{b}\cdot\nabla B\right]Z_i n \rhoe\right)\nonumber\\
-\partial_{w_\parallel}\left(\left[w_\parallel \bm{b}\bm{b}:\nabla\frac{\bm{L}}{\meff n}\right]Z_i n\rhoe\right)=0,\label{eq:kmhd_electron_dke}
\end{gather}
where $m_i$ is the ion mass, $m_e$ is the electron mass, $Z_i$ is the ion atomic number, $\meff=m_i(1+\nu Z_i)$ with $\nu=m_e/m_i$ is an effective mass,
\begin{align}
\mathbb{S}_M=\mu_0^{-1}\frac{B^2}{2}\bm{b}\bm{b}-\mu_0^{-1}\frac{B^2}{2}(1-\bm{b}\bm{b})
\end{align}
is the magnetic part of the Maxwell stress tensor, and 
\begin{align}
\mathbb{P}_i=&m_i n\langle w_\parallel^2\rangle_i\,\bm{b}\bm{b}+n\langle\mu B\rangle_i\,(1-\bm{b}\bm{b}),\\
\mathbb{P}_e=&m_e Z_i n\langle w_\parallel^2\rangle_e\,\bm{b}\bm{b}+Z_i n\langle\mu B\rangle_e\,(1-\bm{b}\bm{b}),
\end{align}
are the ion and electron pressure tensors. Here the angle brackets denote peculiar velocity space averages, e.g. $\langle \mathcal{O}\rangle_i=\int \mathcal{O}\,\rhoi\,dw_\parallel\,d\mu$. 

The interpretation of the dependent variables appearing in this system of equations is the following. The ion number density is $n$, $\bm{L}$ is the (total) momentum density, $\bm{B}$ is the magnetic field, and $\rhoi,\rhoe$ are the \emph{centered conditional probability distribution functions (PDFs)} for ions and electrons, respectively. The centered conditional PDFs are defined such that the probability of finding an ion (electron) in the peculiar velocity-space cell $dw_\parallel\,d\mu$, \emph{given} that the configuration-space location of the ion (electron) is $\bm{x}$, is $\rhoi(\bm{x},w_\parallel,\mu)\,dw_\parallel\,d\mu$ ($\rhoe(\bm{x},w_\parallel,\mu)\,dw_\parallel\,d\mu$.) As such, the centered PDFs are defined to be strictly non-negative and to satisfy the constraints
\begin{gather}
\langle 1\rangle_i=\langle 1\rangle_e=1\label{eq:density_constraint}\\
\langle w_\parallel \rangle_i=\langle w_\parallel\rangle_e=0.\label{eq:velocity_constraint}
\end{gather}
We note Eqs.\,\eqref{eq:density_constraint}-\eqref{eq:velocity_constraint}  are automatically satisfied by solutions of the kinetic equations \eqref{eq:kmhd_ion_dke}-\eqref{eq:kmhd_electron_dke} provided they are satisfied by the initial data.
For the sake of comparing carefully with the equations of Ramos,\cite{Ramos_stab_2016} we also note that the relationship between the centered conditional PDFs and the centered distribution functions $f_i,f_e$ used by Ramos is given by $n\rhoi=\frac{2\pi B}{m_i}f_i$ and $Z_i n\rhoe=\frac{2\pi B}{m_e}f_e$. Finally, it may be interesting to point out that the only essential difference between the form of the GCP equations given here and the form used in Ref.\,\onlinecite{Ramos_stab_2016} is that our formulation uses the total momentum density as a dependent variable instead of the ion fluid velocity. While this is mainly a superficial difference in the context of the present Letter, we have chosen total momentum over velocity with an eye toward future studies of finite-Larmor-radius corrections to the GCP model. As discussed in Ref.\,\onlinecite{Catto_2008} in the different context of low-flow gyrokinetic theory, using $\bm{L}$ as a dependent variable allows one to ``save an order'' in perturbation theory when deriving higher-order variants of GCP. Thus, our use of $\bm{L}$ in the Hamiltonian formulation that we are about to discuss hints at the possibility of using this $\bm{L}$-trick within variational and Hamiltonian formulations of drift-reduced kinetic models, and more generally gyrokinetics.

It follows from the preceding remarks that the guiding center plasma model may be regarded as a first-order ordinary differential equation (ODE) on the infinite-dimensional phase space
\begin{align}
\mathcal{P}=\{&(n,\bm{L},\bm{B},\rhoi,\rhoe)\mid\forall (\bm{x},w_\parallel,\mu)\in Q\times\mathbb{R}\times\mathbb{R}_+\nonumber\\
& n(\bm{x}) > 0\,,\,\rhoi(\bm{x},w_\parallel,\mu)>0\,,\rhoe(\bm{x},w_\parallel,\mu)>0\,,\nonumber\\
&\bm{L}(\bm{x})\in\mathbb{R}^3\,,\,\bm{B}(\bm{x})\in\mathbb{R}^3,\nabla\cdot\bm{B}=0,\nonumber\\
& \langle 1\rangle_i=\langle1\rangle_e=1\,, \langle w_\parallel\rangle_i=\langle w_\parallel\rangle_e=0 \},
\end{align}
where $Q$ is the spatial domain. Here we will assume that $Q$ is either infinite or periodic, and we will refrain from being more specific about the regularity assumptions that should technically be built into $\mathcal{P}$. (For instance we ought to require, at a minimum, and amongst other things, that $n\in L^1(Q)$.) 

The key result of the present work is that the guiding center plasma model, when regarded as a first-order ODE on $\mathcal{P}$, is a noncanonical Hamiltonian system.\cite{Morrison_MHD_1980,Morrison_1980,Marsden_1982} As one might expect, the Hamiltonian is merely the guiding center plasma total energy invariant:
\begin{align}
\mathcal{H}(\mathcal{Z})=&\sum_\sigma\int\left(\langle\mu B\rangle_\sigma+\frac{1}{2}m_\sigma\,\langle w_\parallel^2\rangle_\sigma\right)\,n_\sigma\,d^3\bm{x}\nonumber\\
&+\frac{1}{2}\int \frac{L^2}{\meff n}\,d^3\bm{x}+\frac{1}{2}\mu_0^{-1}\int B^2\,d^3\bm{x}.
\end{align}
where $n_i=n$, $n_e=Z_i n$, and $\mathcal{Z}\in\mathcal{P}$ denotes an arbitrary point in the phase space $\mathcal{P}$. This much could be guessed based on previous investigations of the guiding center plasma model. On the other hand, the model's bracket is by no means obvious. By applying formally-rigorous slow manifold reduction\cite{Burby_two_fluid_2017} to the Vlasov-Maxwell system and exploiting a simple fact about limits of Poisson brackets, we have discovered that the Poisson bracket of arbitrary functionals $\mathcal{F},\mathcal{G}:\mathcal{P}\rightarrow\mathbb{R}$ on the guiding center plasma phase space is given by
\begin{widetext}
\begin{gather}
\{\mathcal{F},\mathcal{G}\}=\int\left[\nabla\times\left(\FBB\right)\cdot\GLL\times\bm{B}-\nabla\times\left(\GBB\right)\cdot\FLL\times\bm{B}\right]\,d^3\bm{x}+\int\left[ \nabla\times\left(\frac{\bm{L}}{n}\right)\cdot\FLL\times\GLL\right]\,n\,d^3\bm{x}\nonumber\\
+\int\left[n\nabla\Fn+n\nabla\left(\FLL\cdot\frac{\bm{L}}{n}\right)\right]\cdot \GLL\,d^3\bm{x}-\int\left[n\nabla\Gn+n\nabla\left(\GLL\cdot\frac{\bm{L}}{n}\right)\right]\cdot \FLL\,d^3\bm{x}\nonumber\\
+\sum_{\sigma}\int\left[\nabla\cdot\left(\left\langle w_\parallel\partial_{w_\parallel}\left(\Frhos\right)\right\rangle_\sigma\,\bm{b}\bm{b}\right)+n\left\langle \nabla\frac{1}{n}\Frhos\right\rangle_\sigma-n\nabla\left\langle\frac{1}{n}\Frhos\right\rangle_\sigma\right]\cdot\GLL\,d^3\bm{x}\nonumber\\
-\sum_{\sigma}\int\left[\nabla\cdot\left(\left\langle w_\parallel\partial_{w_\parallel}\left(\Grhos\right)\right\rangle_\sigma\,\bm{b}\bm{b}\right)+n\left\langle \nabla\frac{1}{n}\Grhos\right\rangle_\sigma-n\nabla\left\langle\frac{1}{n}\Grhos\right\rangle_\sigma\right]\cdot\FLL\,d^3\bm{x}\nonumber\\
+\sum_\sigma\int \bm{b}\cdot\left[\left(\nabla\frac{1}{n_\sigma}\Frhos\right)_{\delta\sigma}-\frac{1}{n_\sigma}\nabla\cdot\left(\left\langle w_\parallel\partial_{w_\parallel}\left(\Frhos\right)\right\rangle_\sigma\,\bm{b}\bm{b}\right)\right]\frac{\partial_{w_\parallel}}{m_\sigma}\left(\frac{1}{n_\sigma}\Grhos\right)\,n_\sigma\overline{\varrho}_\sigma\,d^5\bm{z}\nonumber\\
-\sum_\sigma\int \bm{b}\cdot\left[\left(\nabla\frac{1}{n_\sigma}\Grhos\right)_{\delta\sigma}-\frac{1}{n_\sigma}\nabla\cdot\left(\left\langle w_\parallel\partial_{w_\parallel}\left(\Grhos\right)\right\rangle_\sigma\,\bm{b}\bm{b}\right)\right]\frac{\partial_{w_\parallel}}{m_\sigma}\left(\frac{1}{n_\sigma}\Frhos\right)\,n_\sigma\overline{\varrho}_\sigma\,d^5\bm{z},\label{eq:bracket}
\end{gather}
\end{widetext}
where $d^5\bm{z}=d^3\bm{x}\,dw_\parallel\,d\mu$ and $\mathcal{O}_{\delta\sigma}=\mathcal{O}-\langle \mathcal{O}\rangle_\sigma$ denotes the velocity space fluctuation operator. The remainder of this Letter will be devoted to sketching a derivation of Eq.\,\eqref{eq:bracket}. 

\section{Derivation of the guiding center plasma bracket}
Our derivation of the Poisson bracket \eqref{eq:bracket} starts with the phase space Lagrangian\cite{Cary_1983,Burby_thesis_2015}
\begin{gather}
\mathfrak{L}(Y,\dot{Y})=-\mathcal{G}(\mathcal{Z})\nonumber\\
+\sum_\sigma \int\left(\dot{\psi}_\sigma+\mathcal{L}_\sigma\psi_\sigma\right)\,n_\sigma\,\overline{\varrho}_\sigma\,d^5\bm{z}\nonumber\\
+\sum_\sigma\int q_\sigma \bm{A}_\sigma^*\cdot \bm{u}_\sigma\, n_\sigma\,\overline{\varrho}_\sigma\,d^5\bm{z}\label{eq:PSL}
\end{gather}
where the effective vector potential $\bm{A}_\sigma^*=\bm{A}+\frac{m_\sigma}{q_\sigma}w_\parallel\bm{b}+\frac{m_\sigma}{q_\sigma}\frac{\bm{L}}{\meff n}$, the scalar advection operator $\mathcal{L}_\sigma=\bm{u}_\sigma\cdot\nabla+a_{\parallel\sigma}\partial_{w_\parallel}$, the infinite-dimensional phase space variable $Y=(\bm{z}_i,\bm{z}_e,n,\bm{L},\bm{A},\rhoi,\rhoe,\psi_i,\psi_e)$, $\psi_\sigma$ is a scalar on $(\bm{x},w_\parallel,\mu)$-space, and the Eulerian phase-space fluid velocity $(\bm{u}_\sigma,a_{\parallel\sigma})$ is related to the time derivative of the phase-space fluid configuration map\cite{Holm_1998} $\bm{z}_\sigma(\bm{z}_0)=(\bm{x}_\sigma(\bm{z}_0),w_{\parallel\sigma}(\bm{z}_0),\mu_0)$ according to $\dot{\bm{x}}_\sigma(\bm{z}_0)=\bm{u}_\sigma(\bm{z}_\sigma(\bm{z}_0))$ and $\dot{w}_{\parallel\sigma}(\bm{z}_0)=a_{\parallel\sigma}(\bm{z}_\sigma(\bm{z}_0))$. For the purpose of our derivation, it is important that $\mathcal{G}$ is an arbitrary functional of $\mathcal{Z}=(n,\bm{L},\bm{B},\rhoi,\rhoe)$. The relationship between this Lagrangian and the guiding center plasma model will become clear by the end of the derivation.

The variational principle associated with $\mathfrak{L}$ is $\delta\int \mathfrak{L}\,dt=0$, where $\delta$ indicates arbitrary variations of the phase space trajectory $t\mapsto Y(t)$ with fixed endpoints. Note that $\bm{B}$, $\bm{u}_\sigma$, and $\bm{a}_{\parallel\sigma}$ are not varied directly; their variations are induced\cite{Holm_1998} by variations of $\bm{A}$ and $\bm{z}_\sigma$. Also note that variations of $\overline{\varrho}_\sigma$ are not arbitrary because they must be consistent with the constraints $\langle1\rangle_\sigma=1$ and $\langle w_\parallel\rangle_\sigma=0$. We have computed the variations of the action and found that the Euler-Lagrange equations are given by
\begin{gather}
\dot{\psi}_\sigma+\mathcal{L}_\sigma\psi_\sigma+q_\sigma\bm{A}_\sigma^*\cdot\bm{u}_\sigma\nonumber\\
=\frac{1}{n_\sigma}\Grhos+C_{0\sigma}+w_\parallel C_{1\sigma},\label{eq:delta_rhoi}\\
\partial_t(n_\sigma\overline{\varrho}_\sigma)+\nabla\cdot(\bm{u}_\sigma\,n_\sigma\overline{\varrho}_\sigma)+\partial_{w_\parallel}(a_{\parallel\sigma}\,n_\sigma\,\overline{\varrho}_\sigma)=0\\
q_\sigma \partial_t\bm{A}_\sigma^*+q_\sigma\bm{B}_\sigma^*\times\bm{u}_\sigma+m_\sigma a_{\parallel\sigma}\bm{b}=\nonumber\\
-\nabla\left(\dot{\psi}_\sigma+\mathcal{L}_\sigma\psi_\sigma+q_\sigma\bm{A}_\sigma^*\cdot\bm{u}_\sigma\right)\\
m_\sigma\bm{b}\cdot\bm{u}_\sigma=\partial_{w_\parallel}\left(\dot{\psi}_\sigma+\mathcal{L}_\sigma\psi_\sigma+q_\sigma\bm{A}_\sigma^*\cdot\bm{u}_\sigma\right)\\
\sum_\sigma n_\sigma\langle q_\sigma\bm{A}_\sigma^*\cdot\bm{u}_\sigma\rangle_\sigma+\sum_\sigma n_\sigma\langle\dot{\psi}_\sigma+\mathcal{L}_\sigma\psi_\sigma\rangle_\sigma\nonumber\\
=n\Gn+\frac{1}{1+\nu Z_i}\bm{L}\cdot\langle\bm{u}_i\rangle_i+\frac{\nu Z_i}{1+\nu Z_i}\bm{L}\cdot\langle\bm{u}_e\rangle_e\\
\sum_\sigma q_\sigma n_\sigma\langle\bm{u}_\sigma\rangle_\sigma+\nabla\times\sum_\sigma\left(\frac{m_\sigma n_\sigma}{B}\langle w_\parallel\bm{u}_{\sigma\perp}\rangle_\sigma\right)\nonumber\\
=\nabla\times\GBB\\
\sum_\sigma m_\sigma n_\sigma \langle\bm{u}_\sigma\rangle_\sigma=\meff n \GLL\label{eq:delta_L}
\end{gather}
where $C_{0\sigma}=C_{0\sigma}(\bm{x})$ and $C_{1\sigma}=C_{1\sigma}(\bm{x})$ are unknown functions of $\bm{x}$ that are determined self-consistently by the Euler-Lagrange equations. (The appearance of the $C_{k\sigma}$ is a direct consequence of the constraints $\langle1\rangle_\sigma=1$ and $\langle w_\parallel\rangle_\sigma=0$.) Equations \eqref{eq:delta_rhoi}-\eqref{eq:delta_L} correspond to variations of $\overline{\varrho}_\sigma$, $\psi_\sigma$, $\bm{z}_\sigma$, $n$, $\bm{A}$, and $\bm{L}$, respectively. 

The Euler-Lagrange equations may be used to find an expression for the time derivative $\dot{\mathcal{Z}}_\epsilon=(\dot{n}_\epsilon,\dot{\bm{L}}_\epsilon,\dot{\bm{B}}_\epsilon,\dot{\rhoi}_\epsilon,\dot{\rhoe}_\epsilon)$ in the form of a series in powers of $\epsilon=1/q_i$. (We assume $Z_i=-q_i/q_e=O(1)$.) This is true in spite of the somewhat surprising fact that closed-form expressions for $\dot{\mathcal{Z}}_\epsilon$ valid to all orders in $\epsilon$ are impossible to obtain; the issue is a need to invert nontrivial near-identity first-order differential operators. To leading order, we have found
\begin{gather}
\partial_tn+\nabla\cdot\left(n\GLL\right)=0\\
\partial_t\bm{B}=\nabla\times\left(\GLL\times\bm{B}\right)\\
\meff n\left(\partial_t\frac{\bm{L}}{\meff n}+\nabla\times\left(\frac{\bm{L}}{\meff n}\right)\times\GLL+\nabla\left(\frac{\bm{L}}{\meff n}\cdot\GLL\right)\right)\nonumber\\
=-\sum_\sigma\nabla\cdot\left(\left\langle w_\parallel\partial_{w_\parallel}\Grhos\right\rangle_\sigma\,\bm{b}\bm{b}\right)\nonumber\\
+\sum_\sigma n\nabla\left\langle\frac{1}{n}\Grhos\right\rangle_\sigma-\sum_\sigma n\left\langle\nabla\frac{1}{n}\Grhos\right\rangle\nonumber\\
-n\nabla\Gn+\nabla\times\left(\GBB\right)\times\bm{B}\\
\partial_t(n_\sigma\overline{\varrho}_\sigma)+\nabla\cdot(\bm{u}_\sigma\,n_\sigma\overline{\varrho}_\sigma)+\partial_{w_\parallel}(a_{\parallel\sigma}\,n_\sigma\overline{\varrho}_\sigma)=0,
\end{gather}
where
\begin{align}
\bm{u}_\sigma&=\GLL+\left(\frac{\partial_{w_\parallel}}{m_\sigma}\frac{1}{n_\sigma}\Grhos\right)_{\delta\sigma}\bm{b}\\
a_{\parallel\sigma}&=\bm{b}\cdot\bigg[\frac{1}{m_\sigma n_\sigma}\nabla\cdot\left(\left\langle w_\parallel\partial_{w_\parallel}\Grhos\right\rangle_\sigma\,\bm{b}\bm{b}\right)\nonumber\\
&\hspace{1.5em}-\left(\nabla\frac{1}{m_\sigma n_\sigma}\Grhos\right)_{\delta\sigma}+w_\parallel\nabla\left\langle\frac{\partial_{w_\parallel}}{m_\sigma }\frac{1}{n_\sigma}\Grhos\right\rangle\nonumber\\
&\hspace{9em}-w_\parallel\bm{b}\cdot\nabla\GLL\bigg].
\end{align}

The ($\epsilon$-dependent) Poisson bracket associated with the phase space Lagrangian \eqref{eq:PSL} is determined by equating the time derivative of an arbitrary observable $\mathcal{F}$ along the dynamics generated by the Hamiltonian $\mathcal{G}$ with (a) the usual expression from Hamiltonian mechanics $\dot{\mathcal{F}}_\epsilon=\{\mathcal{F},\mathcal{G}\}_\epsilon$ and (b) the expression for the time derivative given by the functional chain rule. In particular,
\begin{align}
\{\mathcal{F},\mathcal{G}\}_\epsilon&=\int \dot{n}_\epsilon\Fn\,d^3\bm{x}+\int\dot{\bm{L}}_\epsilon\cdot\FLL\,d^3\bm{x}\nonumber\\
&+\int\dot{\bm{B}}_\epsilon\cdot\FBB\,d^3\bm{x}+\sum_\sigma\int\dot{\overline{\varrho}}_{\sigma\epsilon}\Frhos\,d^5\bm{z},
\end{align}
which satisfies the Jacobi identity automatically as a result of being derived by inverting the symplectic form associated with a phase space Lagrangian. (See Ref.\,\onlinecite{Burby_thesis_2015} for a variety of example applications of this procedure.) Moreover, because we have shown that $\dot{\mathcal{Z}}_\epsilon=O(1)$ has a well-defined limit as $\epsilon\rightarrow 0$, it follows that $\dot{\mathcal{F}}_\epsilon=O(1)$ has a well-defined limit for any observable $\mathcal{F}$. Therefore the limiting bracket $\{\mathcal{F},\mathcal{G}\}_0=\lim_{\epsilon\rightarrow 0}\{\mathcal{F},\mathcal{G}\}_\epsilon=\dot{\mathcal{F}}_0$ is well defined, and we deduce the simple consequence
\begin{align}\label{limit_jacobi}
\lim_{\epsilon\rightarrow 0}\circlearrowright\{\{\mathcal{F},\mathcal{G}\}_\epsilon,\mathcal{H}\}_\epsilon=\circlearrowright\{\{\mathcal{F},\mathcal{G}\}_0,\mathcal{H}\}_0,
\end{align}
where $\circlearrowright$ indicates a sum over cyclic permutations of $\mathcal{F}$, $\mathcal{G}$, and $\mathcal{H}$. Equation\,\eqref{limit_jacobi} asserts that the limiting bracket $\{\mathcal{F},\mathcal{G}\}_0$
satisfies the Jacobi identity. Because it is straightforward to show that $\{\mathcal{F},\mathcal{G}\}_0$ reproduces Eq.\, \eqref{eq:bracket}, this proves that the bracket \eqref{eq:bracket} satisfies the Jacobi identity. Moreover, Hamilton's equations $\dot{\mathcal{F}}=\{\mathcal{F},\mathcal{H}\}_0$ for an arbitrary observable $\mathcal{F}$ may be shown to reproduce the guiding center plasma model. This proves that Eq.\,\eqref{eq:bracket} is indeed the Poisson bracket for the guiding center plasma model.

\section{Discussion}

It is somewhat surprising that the Hamiltonian formulation of the guiding center plasma model has gone unnoticed for so long. This point is underscored by Morrison and Greene's discovery\cite{Morrison_MHD_1980} of the Hamiltonian formulation of ideal MHD in 1980. Perhaps the reason that this gap in our understanding of the GCP model has persisted for as long as it has is the complexity of the bracket \eqref{eq:bracket} relative to the ideal MHD bracket, which is probably suficcient to foil educated guesses. The reason may also be related to the fact that Ramos' formulation\cite{Ramos_KMHD_2015,Ramos_2008} of kinetic MHD in the mean-flow frame, which was immensely helpful in our investigation, was only formulated relatively recently.

While this Letter identifies a previously-unknown structural property of the GCP model, that property is admittedly abstract. Are there more concrete corollaries to the presence of this structure? We have found the following partial answer: the GCP Hamiltonian structure reveals new circulation theorems -- kin to the well-known Kelvin circulation theorem.  We have identified five circulation theorems, four of which appear to be new discoveries. We will conclude this note by asserting these theorems without proof.

The most obvious invariant is the magnetic flux through a loop in configuration space that is passively advected by the fluid velocity $\bm{U}\equiv \bm{L}/\meff n$. The invariance of this quantity is merely a restatement of the frozen-in condition \eqref{eq:kmhd_faraday}, and is therefore well-known. There are two other circulation invariants that are related to, but not equivalent to the configuration-space magnetic flux invariant. Specifically, the magnetic flux through a loop in \emph{phase space} that is passively advected by the phase space flow of either electrons or ions is conserved. Lastly, there are two circulation invariants that are related to the second adiabatic invariant. Suppose that a constant-$\mu$ loop in phase space has a configuration-space projection that lies along (part of) a single magnetic field line. If that loop is passively advected by either the electron or ion phase space flow, then the integral $\oint w_\parallel\bm{b}\cdot d\bm{x}$ is constant in time. The condition on the loop's configuration-space projection is essential here; if the projection intersects more than one field line, then the circulation is not generally conserved. When proving this circulation theorem, it is necessary to make use of the following topological feature of the GCP model. An advected phase space loop that projects onto a single field line at one instant of time also projects onto a single (but generally different) field line at all other times.


The work presented here has benefited from various helpful discussions with Harold Weitzner, Antoine Cerfon, Jesus Ramos, Greg Hammett, and Alain Brizard. This research was supported by the U. S. Department of Energy, Office of Science, Fusion Energy Sciences under  Award No. DE-FG02-86ER53223  and the U.S. Department of Energy Fusion Energy Sciences Postdoctoral Research Program administered by the Oak Ridge Institute for Science and Education (ORISE) for the DOE. ORISE is managed by Oak Ridge Associated Universities (ORAU) under DOE contract number DE-AC05-06OR23100. All opinions expressed in this paper are the author's and do not necessarily reflect the policies and views of DOE, ORAU, or ORISE. This work was also supported by the U.S. Department of Energy Grant No. DEFG02-86ER53223.


\providecommand{\noopsort}[1]{}\providecommand{\singleletter}[1]{#1}%
%


\end{document}